\documentclass[lettersize,journal]{IEEEtran}
\usepackage{diagbox}
\usepackage{subcaption}
\usepackage[table]{xcolor}
\usepackage{amsmath,amsfonts}
\usepackage{algorithm}
\usepackage{algpseudocode}
\usepackage{multirow}
\usepackage{array}
\definecolor{Gray}{gray}{0.85}
\usepackage{threeparttable}
\usepackage[caption=false,font=normalsize,labelfont=sf,textfont=sf]{subfig}
\usepackage{textcomp}
\usepackage{stfloats}
\usepackage{url}
\usepackage{verbatim}
\usepackage{graphicx}
\def\BibTeX{{\rm B\kern-.05em{\sc i\kern-.025em b}\kern-.08em
    T\kern-.1667em\lower.7ex\hbox{E}\kern-.125emX}}
\usepackage{balance}
\begin{document}
\title{Federated Sequence-to-Sequence Learning for Load Disaggregation from Unbalanced Low-Resolution Smart Meter Data}

\author{Xiangrui~Li
\thanks{X. Li is with the Department of Data Science and AI, Faculty of Information Technology and Monash Energy Institute, Monash University, Clayton, VIC 3800, Australia (e-mails: xlii0281@student.monash.edu.}
}



\maketitle

\begin{abstract}
The importance of Non-Intrusive Load Monitoring (NILM) has been increasingly recognized, given that NILM can enhance energy awareness and provide valuable insights for energy program design.
Many existing NILM methods often rely on specialized devices to retrieve high-sampling complex signal data and focus on the high consumption appliances, hindering their applicability in real-world applications, especially when smart meters only provide low-resolution active power readings for households. In this paper, we propose a new approach using easily accessible weather data to achieve load disaggregation for a total of 12 appliances, encompassing both high and low consumption, in scenarios with very low sampling rates (hourly). Moreover, We develop a federated learning (FL) model that builds upon a sequence-to-sequence model to fulfil load disaggregation without data sharing. Our experiments demonstrate that the FL framework - L2GD can effectively handle statistical heterogeneity and avoid overfitting problems. By incorporating weather data, our approach significantly improves the performance of NILM.
\end{abstract}

\begin{IEEEkeywords}
Non-Intrusive Load Monitoring (NILM), load disaggregation, sequence-to-sequence, federated learning.
\end{IEEEkeywords}

\section{Introduction}
\IEEEPARstart{R}{ecent} years have seen a rise in the adoption of smart meters in households globally, primarily driven by energy-saving targets. These devices have the potential to increase energy awareness, thereby reducing power use \cite{benzi2011electricity}. However, most smart meters only record the total household consumption, making it necessary to disaggregate energy consumption into appliance levels. The non-intrusive load monitor (NILM) method proposed by Hart \cite{hart1992nonintrusive} offers a cost-effective and convenient way to disaggregate total energy consumption into individual appliance consumption, thereby improving energy awareness and efficiency. The development of accurate NILM methods holds theoretical and practical significance for energy consumers, utilities, and policymakers. Apart from identifying inefficient appliances, a detailed breakdown of electrical consumption at the appliance level would allow residences to replace energy-inefficient appliances with more efficient ones, resulting in significant energy savings.

NILM initially aims to leverage signal characteristics, such as harmonics, current, and voltage, to identify appliances' status and disaggregate total energy consumption into individual appliances' levels. In the studies of the last two decades, event-based methods NILM approaches are widely used, such as \cite{eb2018,eb2020,7007702,stankovic2016measuring}. This approach often requires dedicated devices with a high sampling rate (typically \textless 1 second) to detect distinctive features or changes in the electrical signal that indicate the activation or deactivation of specific appliances. These events could be transient changes, patterns, or signatures in the power signal that are characteristic of certain appliances. In other studies, some deep learning methods have been utilized to identify the switch events in the high sampling scenerios. Such as convolutional neural network (CNN)-based methods\cite{CNNmulti,zhou2020sequence,CNNBilstm,kelly2015neural} and denoising autoencoder (DAE) \cite{wang2019new,kelly2015neural}. Different from the event-based methods that identify the events through manually selected thresholds or rules, these methods enable the extraction of relevant features related to event switch changing from complex, high-sampling data by adjusting parameters through training.

Real-world scenarios, however, pose challenges as smart meters in general households are unable to capture complex signals, such as harmonics, transients, or current-voltage signals, but only active power readings. Furthermore, residential smart meters were originally installed to record household electricity consumption for billing purposes. Thus, the sampling rate of smart meters is typically rated from 15 minutes to 1 hour. According to \cite{basu2015time} and \cite{7793294}, considering the limitation of the sampling rate of household electricity meters, event-based methods require a high sampling rate for event switching detection are not appropriate. The main obstacle lies in detecting low-energy-consuming devices because their changes in active power are not significant compared to high-power appliances.  The data from household electricity meters, being averages of loads over 15-minute to 1-hour time intervals, exhibit more pronounced changes in the load of high-consumption appliances, while the load changes of low-consumption appliances are smoothed out due to averaging, making challenges to detect their switching events. For the CNN and DAE methods, compared to dedicated devices, household electricity meters have lower sampling frequencies and therefore collect less data. This increases the training difficulty for methods like CNN and DAE, which require large amounts of data. Additionally, because these methods lack temporal dependency and treat each time step equally, they face similar challenges to event-based methods in detecting low-energy-consuming appliances in the averaged meter records. Some other works \cite{6915905,basu2015time}
are only focused on the high-consuming appliances, such as water heaters and air conditions ignoring the low energy-consuming devices. However, in the context of real load disaggregation tasks, the consumption of low-power appliances is equally crucial. This is because the usage patterns of these appliances are often associated with the residents' electricity habits. 

Considering the limitations of sampling rate in household electric meters, some studies suggest that the accuracy of NILM algorithms can be improved by enhancing the diversity of data or augmenting data from household electricity meters. In the work \cite{quy2021data,fagiani2019non}, the data argument method was utilized to generate high sampling rate data from low rate
samples by using step-wise interpolation. Generating the high sampling rate data, the methods used in the high sampling data can be adopted. However, interpolation typically fills in missing data points based solely on the observed values before and after the gap, without considering the broader context of the data and may lead to a loss of information and potentially distort the original energy consumption pattern. Different from the data argument, some studies have explored the use of weather information to better understand energy consumption behaviours. In \cite{kipping2016modeling,birt2012disaggregating}, both methods analyzed hourly energy readings collected from households to gain insights into energy consumption patterns and disaggregate them into specific end-uses.
In the work of \cite{zhang2015residential}, they introduced a multi-objective genetic algorithm alongside pre-learned deductions about household appliances. These deductions are based on previous observations of both active and reactive energy usage, weather conditions, and details regarding appliance ownership.
Compared with the complex signals, the weather data is readily accessible and has been validated in previous works as a useful feature.
 
Another challenge is caused by data privacy, due to the potential risk of data leakage. Users' electricity usage patterns and personal information can be inferred if their meter data are leaked. Therefore, under strict data protection regulations, the accessibility of meter data faces challenges, thereby increasing the difficulty of model training. Limited data also raises the risk of overfitting during model training. To ensure reliability in training models while also fulfilling privacy concerns, the Federated learning (FL) framework has been leveraged in the NILM tasks. In the work of \cite{giuseppi2022decentralized,kaspour2022federated}, FedAvg has been used to enable the training of the model without data sharing. However, FedAvg will encounter the issue of slow convergence when data is heterogeneous between clients. In real-world scenarios, the diverse usage habits of residents regarding electrical appliances, along with variations in weather conditions, pose challenges in dealing with data heterogeneous. To mitigate the influence of data heterogeneous, FedProx \cite{li2020federated} is generally used in the data heterogeneous scenarios. In the NILM tasks, however, both the FedAvg and FedProx require a significant amount of communication between the server and clients, which poses a significant challenge in terms of communication and computational overhead for NILM devices. Considering the limitations in computational efficiency and computational costs of NILM devices, additional methods are needed to strike a trade-off between communication frequency and model accuracy. 

Inspired by the approaches listed above, to tackle the challenges that are caused by the limited sampling rate of smart meters and data privacy, this paper proposes a sequence-to-sequence (seq2seq) federated learning model that utilizes an encoder to extract global information from input weather and smart meter load data, providing it as prior input to the decoder that performs hourly load disaggregation without data sharing. 
Specifically, we propose a new sequence-to-sequence model that takes in 24-hour temperature and humidity information in addition to total energy consumption to disaggregate the energy usage of 12 appliances. To ensure no data sharing during the training and avoid model overfitting in the limited data, we implement the model on federated learning frameworks.Furthermore, in order to mitigate communication overhead during Federated Learning (FL) training and address the challenge of heterogeneous data, we deploy the L2GD federated learning framework in Non-Intrusive Load Monitoring (NILM) tasks.
The contributions of this paper can be summarized as follows.
\begin{itemize}
  \item We have proposed a Seq2Seq-based method that utilizes time-series data of total load and weather to disaggregate the total load into 12 appliances with varying power consumption, including low-power devices. Simultaneously, 
  considering in the real scenarios that residential electric usage patterns are likely influenced by the season variations, we sample the data in two strategies to simulate the real scenarios and evaluate the proposed model within three federated learning frameworks.
  \item Considering the difficulty in identifying the low-power appliances in hourly scenarios from total load data, we incorporate temporal weather data intending to identify the load of low-power appliances from the total load by leveraging weather-related information. Moreover, unlike previous approaches, we propose a seq2seq model. The encoder captures global information from the entire time series instead of focusing solely on load changes in specific time slots. This global information is then passed to the decoder as prior knowledge. The decoder integrates both global and individual time step information, enabling the disaggregation of different power appliances in hourly scenarios.
  \item Considering the diverse data distribution among households in real-world scenarios and the significant communication overhead associated with federated learning, we have introduced the L2GD framework for the first time in the NILM domain. This framework aims to reduce communication costs while mitigating the negative impact of data heterogeneity on training. Experimental results indicate that, in both data homogenous and heterogenous scenarios, L2GD achieves comparable performance with approximately half the communication rounds compared to FedAvg and FedProx. Furthermore, in situations with an equal number of communication rounds and under data heterogenous conditions, it demonstrates superior effectiveness.
\end{itemize} 
\section{Related works}
In this section, we will review the related works that use time series deep-learning techniques in the NILM tasks and federated learning frameworks in privacy concern scenarios. 

In deep learning methods, the NILM tasks are usually formulated as time series classification or regression problems. Thus recurrent neural network (RNN) based methods have been widely used in recent years. Diego \textit{et al.} \cite{de2021recurrent} takes two layers stacked LSTM to do the load disaggregation task, which concatenates the hidden states from the last layer with input data and feeds them into the dense layers. Another approach can be found in \cite{mauch2015new,kaselimi2019bayesian}, the bidirectional LSTM (BiLSTM) is taken to retrieve the hidden states from forward and backward directions low-sampling sampling data. Although the BiLSTM can handle long sequences and capture more effective hidden states than stacked LSTM, there is useless information contained in the input sequence. To better extract features from input data, in the work of \cite{wang2019new}, the denoising autoencoder (DAE) is taken to reconstruct the original data that only retains the principle component in the input sequence. Then, the reconstructed sequence is fed input to the LSTM. A similar approach can be found in Kelly \textit{et al.} works \cite{kelly2015neural}, which add several 1D convolutional layers in the DAE and trained by manually corrupting the signal before feeding it into the input layer. According to the experience results, the addition of a convolution layer can slightly increase the performance against the work of \cite{wang2019new}. In the work of \cite{zhou2022non}, the researcher substituted the LSTM as Nest LSTM (NLSTM). Different from conventional LSTM, NLSTM contains an internal and external unit, which collaborates to choose and retain the most important long-term information based on the current situation. It helps create a strong and selective memory of the long-term characteristics of the target device. Although the LSTM can store the context information of time-series data, it has no ability to decide what parts of the input sequence are most important for the output. To mitigate this problem, the attention mechanism is implemented in \cite{zhou2022non,liu2022nonintrusive,fan2022bidirectional,zhang2021improving} to evaluate the similarity between the hidden states of different time steps. The CNN-based methods can be found in \cite{CNNmulti,zhou2020sequence}. The main idea is to project the input sequence to feature maps on a temporal scale. The CNN block is utilized to integrate the more advanced features with the previously calculated high-resolution features.

To achieve user privacy protection and computational efficiency, federated learning (FL) has been used in the field of NILM in recent years. In the study conducted by Hudson \textit{et al.} \cite{hudson2021framework}, the FedAvg is implemented with the recurrent neural network architecture, where smart houses participated in the training process by utilizing their recorded meter data. The model parameters learned through federated learning were shared using the advanced metering infrastructure (AMI), ensuring the privacy of the data. A similar approach can be found in \cite{Seq2point_FL}, the Seq2piont model is combined with the FedAvg framework in the NILM task. However, although the previous studies achieved good performance by utilizing the FL, they have shortages to deal with the domain adaption problem. To mitigate the domain adaption problem, Li \textit{et al.} \cite{li2021energy} utilized transfer learning to incorporate FedAvg to identify and learn individual equipment states retrieved from different domains. Different from \cite{li2021energy}, the meta-learning is leveraged in the work of \cite{liu2022learning}, which trains the meta-model on the FL framework and fine-tunes each client. In the work of \cite{hanzely2020federated}, the L2GD was proposed as to mixture of the global and clients' models. Different from the FedProx \cite{li2020federated}, L2GD can not only trade-off between the global and client model but also can utilize hyperparameters to adjust communication rounds, which can effectively reduce communication costs.

\section{methodology}
\begin{figure*}[ht]
    \centering 
    \includegraphics[width=1.0\linewidth]{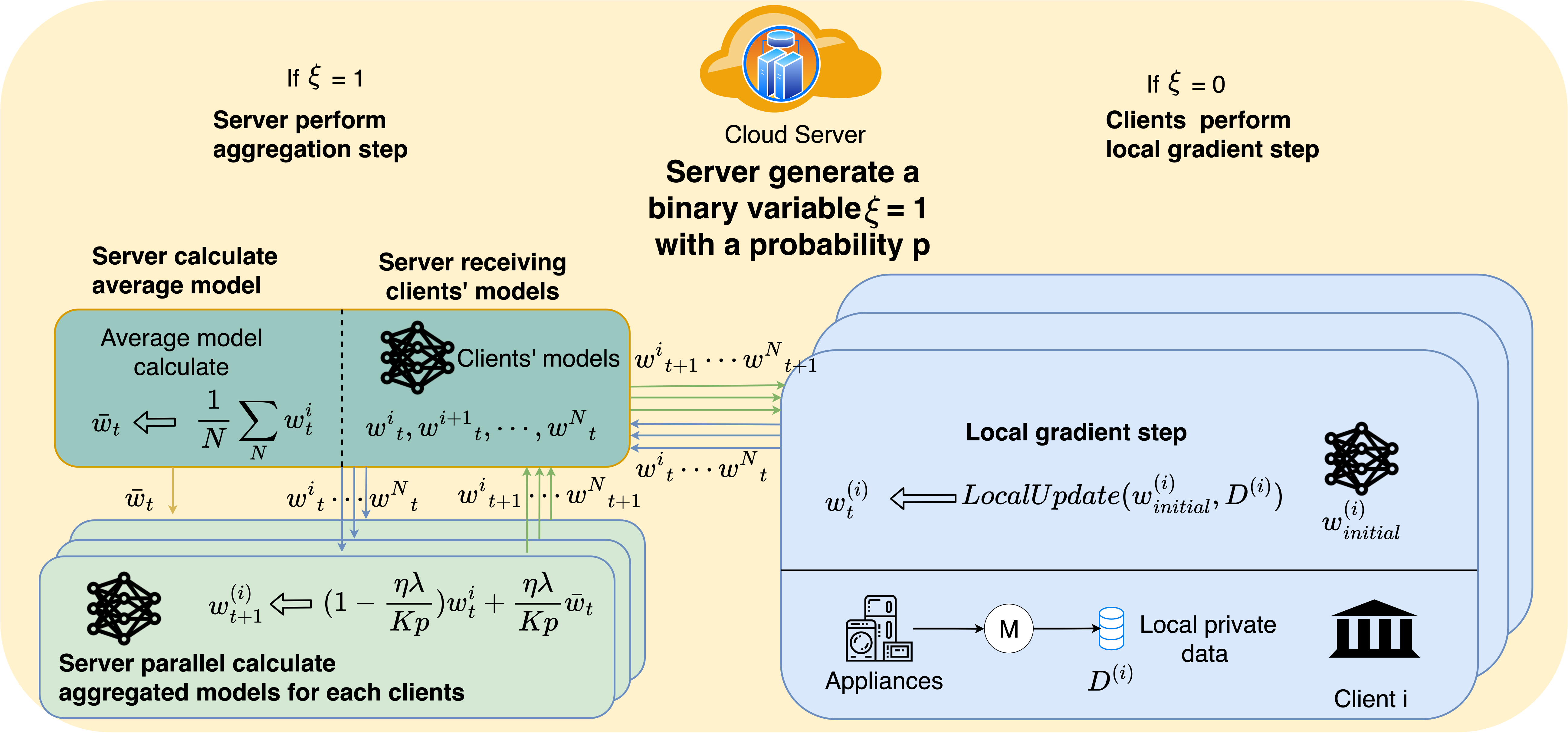} 
    \caption{The overview of L2GD framework.}
    \label{fig:overview_frame}
\end{figure*}

Household electricity meters have the characteristic of low sampling frequency, causing difficulties in extracting patterns of low-power appliances' usage. In this regard, we propose a sequence-to-sequence model that utilizes LSTM with memory and forgetting capabilities to capture contextual information. We incorporate time-series weather data as additional features and leverage the potential relationship between weather changes and appliance usage to simultaneously disaggregate the loads of 12 appliances, including both high and low power, in hourly scenarios. To ensure privacy protection, we integrate the sequence-to-sequence model into the L2GD framework, ensuring that only model parameters are shared during training while users' private data remain secure. The overview framework is illustrated in Fig.\ref{fig:overview_frame}. Subsequently, we will discuss the sequence-to-sequence model and federated learning framework in the following sections.

\subsection{Sequence-to-Sequence Model}
\begin{figure}[t]
    \centering 
    \includegraphics[width=0.8\linewidth]{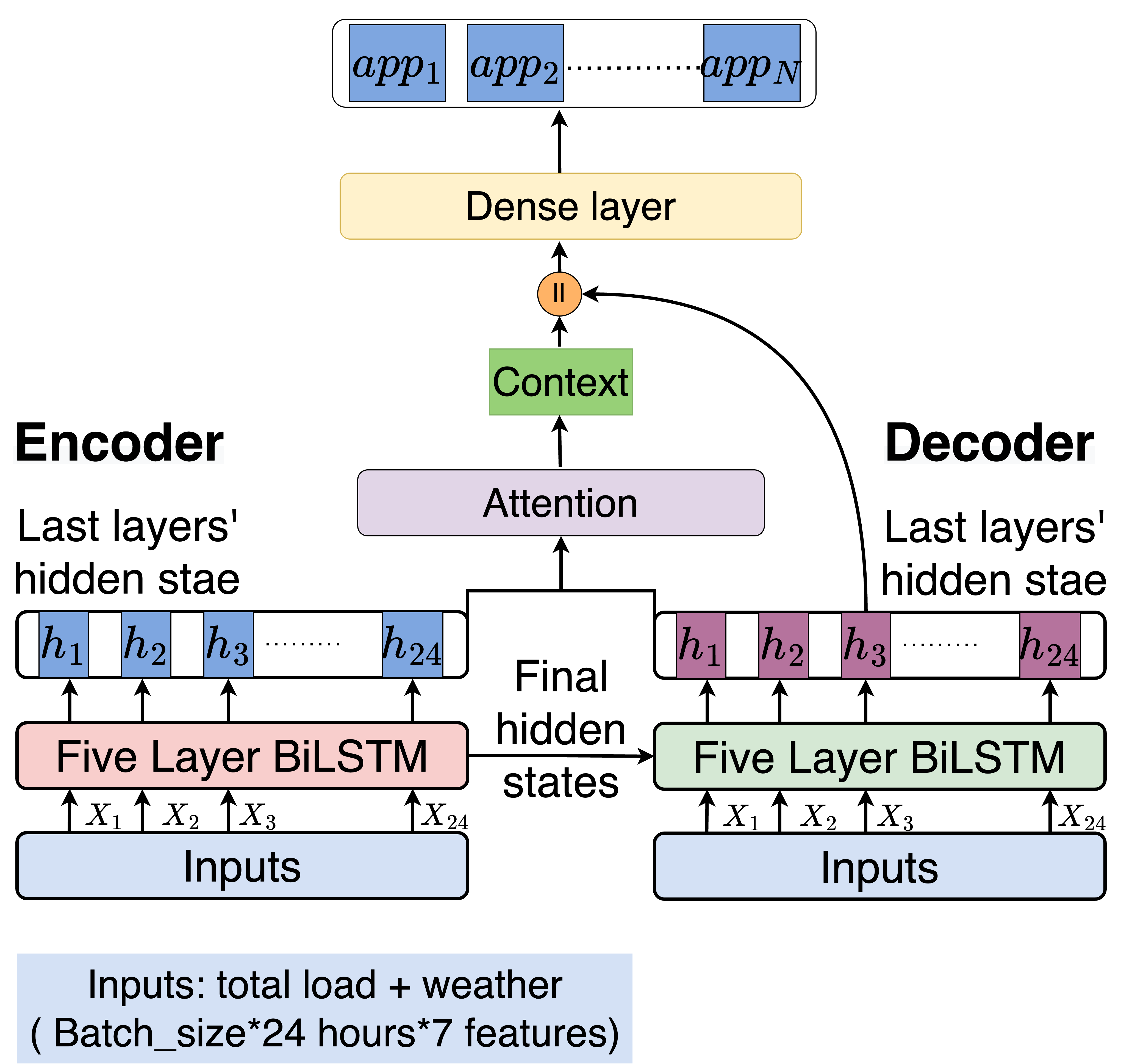} 
    \caption{Proposed Sequence to Sequence model.}
    \label{fig:Seq2Seq_model}
\end{figure}

The Sequence-to-Sequence model is composed of three components, including an encoder, decoder, and attention part. 
Fig.~\ref{fig:Seq2Seq_model} shows the detail regarding the model structure.
The encoder is composed of stacked bidirectional LSTM, which is responsible for encoding the input data and outputting the last time steps' hidden states $H_{lte}$, cell states $C_{lte}$ and final layers hidden states $H_{fle}$. The input data, including total consumption and historical weather information, is fed into the encoder part first. 
\begin{equation}
\begin{aligned}
&H_{fle},H_{lte},C_{lte}=\text{Encoder}(\boldsymbol{X})\\
&\boldsymbol{X} = \text{Concatenate[weather;total load]}
\end{aligned}
\end{equation}

Due to gating mechanisms in LSTM, the $H_{lte}$ and $C_{lte}$ can store the contextual information regarding the whole times-series load and weather-changing and $H_{fle}$ stores the hidden states of each time step. In the decoder part, we use the same structure as the encoder but initial the decoder hidden states and cell states by using the last time step hidden $H_{lte}$ and cell states $C_{lte}$ from the encoder, which stores the contextual information of whole time-series information. Then, the decoder takes the same input data for decoding and outputs the decoded final layers' hidden states $H_{fld}$. Using the hidden states from the last time step of the encoder to initialize the decoder allows the decoder to be more sensitive to changes in the input load at each time step while having knowledge of the global contextual information.
\begin{equation}
\begin{aligned}
H_{fld}=\text{Decoder}(\boldsymbol{X},H_{lte},C_{lt})
\end{aligned}
\end{equation}

By initializing the decoder with the last time step hidden $H_{lte}$ and cell states $C_{lte}$ from the encoder, the decoder updated details in the time step $t_0$ should be:
\begin{equation}\label{eq:first update}
\begin{aligned}
i_0 &= \sigma(W_{ii} x_0 + b_{ii} + W_{hi} H_{lte} + b_{hi}) \\
f_0 &= \sigma(W_{if} x_0 + b_{if} + W_{hf} H_{lte} + b_{hf}) \\
g_0 &= \tanh(W_{ig} x_0 + b_{ig} + W_{hg} H_{lte} + b_{hg}) \\
o_0 &= \sigma(W_{io} x_0 + b_{io} + W_{ho} H_{lte} + b_{ho}) \\
c_0 &= f_0 \odot C_{lte} + i_0 \odot g_0 \\
h_0 &= o_0 \odot \tanh(c_0) \\
\end{aligned}
\end{equation}

\begin{figure}[ht]
    \centering 
    \includegraphics[width=1.0\linewidth]{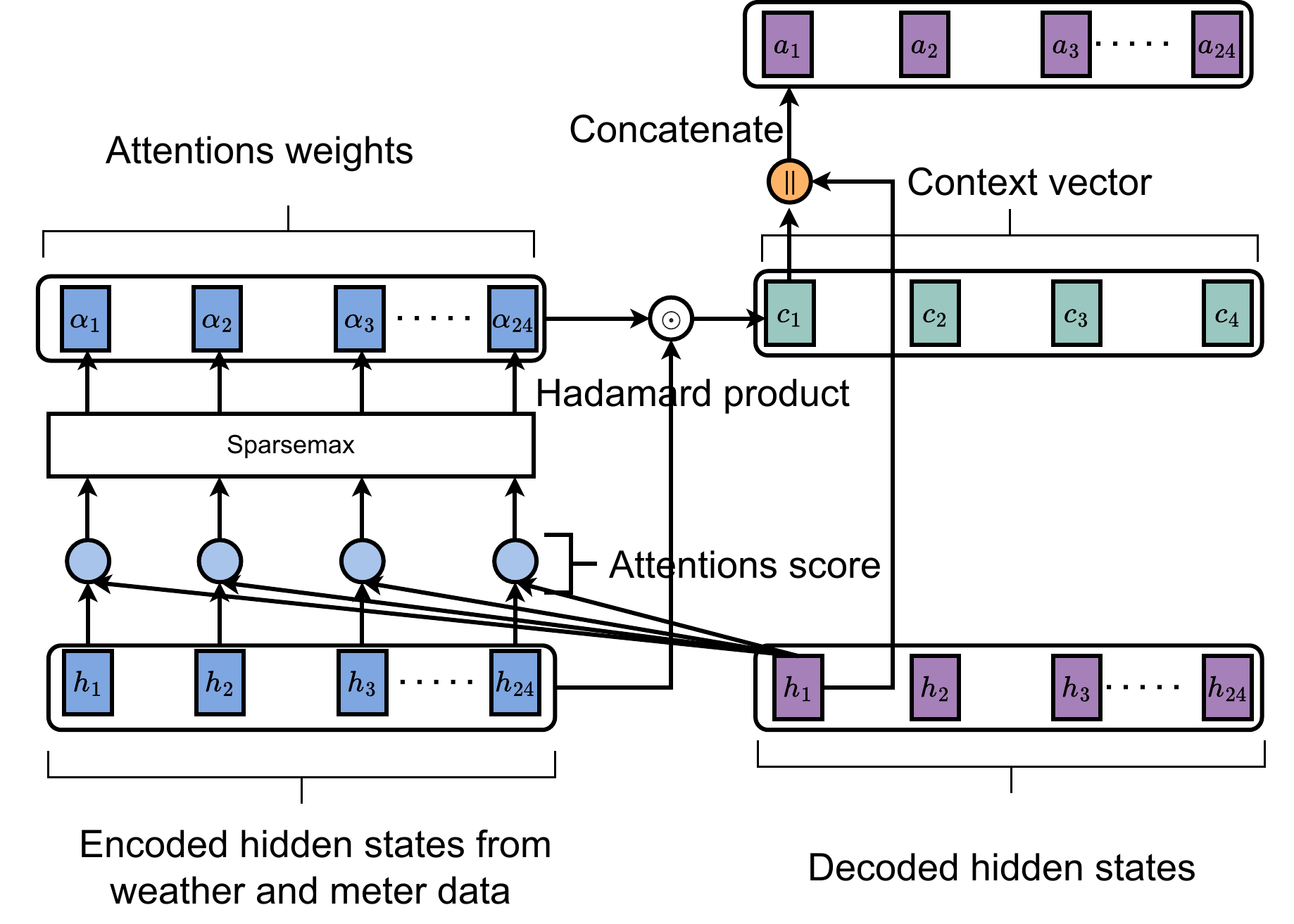} 
    \caption{Attention units implemented in Sequence-to-Sequence model.}
    \label{fig:attention}
\end{figure}

According to Equation \ref{eq:first update}, with each time step update, the $H_{lte}$ and $C_{lte}$, which contain global information used to initialize the decoder, will gradually be forgotten.
To partially retain the global information retrieved by the encoder,  the attention unit is implemented before the dense layer. The structure of the attention unit is shown in Fig \ref{fig:attention}.

We gain insight from the work of \cite{luong-etal-2015-effective} to implement the attention on the output from the encoder and decoder. Denoting $\bar{\boldsymbol{h}}_s$ and $\boldsymbol{h}_t$ as the encoder final layer's hidden states and the decoder final layer's hidden state. All these hidden state outputs contain processed time-series information of the model inputs. The attention unit first evaluates the similarity of hidden states from different time steps using the dot product.
\begin{equation}
    score(\boldsymbol{h}_t, \bar{\boldsymbol{h}}_s)=\boldsymbol{h}_t^{\top} \bar{\boldsymbol{h}}_s,
\end{equation}
Then the attention unit normalizes the score using the normalization function. In the common attention model, the Softmax is used as the normalization function to normalize the attention score between 0 to 1. However, in our task, we want to use attention to calculate the similarity between different time steps of the encoder and decoder. The hidden states at a certain time step t contain information from previous all-time steps. Using Softmax in this task would make the weights assigned to each time step become more evenly distributed. Thus we take the Sparsemax function \cite{Sparsemax} to obtain the attention weight $\boldsymbol{\alpha}_{ts}$ according ${Sparsemax}(score(\boldsymbol{h}_t, \bar{\boldsymbol{h}}_s))$, which is used to generate the context vector $\boldsymbol{c}_t$ in Eq. \eqref{equ:Context Vector}.
\begin{equation}
\begin{aligned}
&\text{Sparsemax}(z)_i = \max(0, z_i - \tau(z)), \\
&\tau(z) = \frac{1}{k}\left(\sum_{i=1}^n z_i - 1\right)
\end{aligned}
\end{equation}
where $ i=1,2,\ldots,n$, and $k$ is the number of non-zero elements in $\operatorname{sparsemax}(z)$. 

After getting the attention weights, the Hadamard product is used to weight the hidden states from the encoder and the output of the attention unit $\boldsymbol{a}_t$, called attention vector, is obtained in Eq. \eqref{equ:Attention Vector}:
\begin{gather}
	\boldsymbol{c}_t = \sum_{s}\boldsymbol{\alpha}_{ts}\bar{\boldsymbol{h}}_s, \label{equ:Context Vector}\\
	\boldsymbol{a}_t = [\boldsymbol{c}_t; \boldsymbol{h}_t], \label{equ:Attention Vector}
\end{gather}
where $[\boldsymbol{c}_t; \boldsymbol{h}_t]$ concatenate $\boldsymbol{c}_t$ and $\boldsymbol{h}_t$ into one matrix. 

The concatenate vectors will be fed into several dense layers to output the disaggregated loads for each appliance.
\subsection{Federated Learning}\label{FL_methodology}
In our NILM task, one of the potential challenges is the statistic heterogeneous problem, which arises due to the diverse electricity usage patterns among different residences. Various factors, such as residents' working hours, socioeconomic status, and household size, can affect appliance usage distribution, resulting in a high variance of customized model parameters and potential overfitting problems. To address this problem, we use two federated learning frameworks (e.g., FedAvg and FedProx) in Algorithm \ref{alg:Alg_Fed} to assess how the statistical heterogeneity impacts model performance and whether the FedProx can effectively handle this challenge. Another challenge is caused by the communication costs, especially for the heterogeneous scenarios. According to the \cite{li2020federated}, the data heterogeneous has an influence on the model convergence. The more significant the difference in data distribution among clients, the more communication requests each client needs to make to the server. In our works, we combine L2GD \cite{hanzely2020federated} with our proposed Seq2Seq model Algorithm \ref{alg:L2GD}, which allows a trade-off between the global model and the local models, to reduce the computation complexity by adjusting the aggregation frequencies. The L2GD framework can be found in Fig.\ref{fig:overview_frame}. When initiating L2GD, the server sends the initial model to each client. Simultaneously, with a probability p, it generates $\xi$. If $\xi$ is 0, each client trains a local model using its private data. If $\xi$ is 1, each client sends its local model to the server. The server then computes the average model and sends it back to the client. The client updates its model based on the dissimilarity between the average model and the local model, as well as the hyperparameter $\lambda$. When $\lambda$ is equal to 0, L2GD is equivalent to local update. As $\lambda$ increases, the client model gradually approaches the average model. 

\begin{algorithm}[ht]
  \caption{FedAvg \& FedProx FL frameworks for our NILM task. (Note that $\frac{\mu}{2}{ \|w^{Global}- w\| }^2$ is only used in FedProx.)}
  \label{alg:Alg_Fed}
  \begin{algorithmic}
    \State \textbf{Require:} Number of household: $K$, number of communication rounds: $T$, number of epoch in each communication rounds: $E$, learning rate: $\eta$, global model initial parameters: $w^{Global}_{0}$
    
    \State \textbf{Initial:} $w^{Global}\leftarrow{w^{Global}_{0}}$
    \For{$t = 0, \dots, T-1$}
      \State Server random selects a subset $S_k$ of $K$ household clients(each clients  is chosen with probability $p$)
      \State Dispatch $w^{Global}$ to each client  $ i \in S_k$
      \For{each client  $ i \in S_k$, client private data $D^{i} $}
         \State  $w^{i}_{t+1}\leftarrow \Call{LocalUpdate}{w^{Global},D^{i}}$
        \State \textbf{Calculate data proportion:} $\gamma^{i} \leftarrow \frac{D^{i}}{D}$
      \EndFor
      
      \State  $w^{Global}\leftarrow \sum_{S_k}\gamma^{i}{w^{i}_{t+1}}$
      \EndFor
    \State \textbf{Return} $w^{Global}$ 
    \\\hrulefill
    \Function{LocalUpdate}{$w^{Global},D^{i}$}:
    \For{epoch $e$ in range($E$)} 
    \State $Loss = L(w;D^{i})$ + $\frac{\mu}{2}{ \|w^{Global}- w\| }^2$
    \State $w^{i}_{t+1} \leftarrow w^{i}_{t} - \eta\nabla Loss$ 
    \EndFor
    \State \textbf{Return} $w^{i}_{t+1}$ 
    \EndFunction

  \end{algorithmic}
\end{algorithm}

\begin{algorithm}[ht]
  \caption{L2GD for NILM tasks}
  \label{alg:L2GD}
  \begin{algorithmic}
    \State \textbf{Require:} Number of household: $K$, number of communication rounds: $T$, number of epoch in each communication rounds: $E$, learning rate: $\eta$, global model initial parameters: $w^{Global}_{0}$, probability to perform aggregation step: $p$, hyperparameter to trade off between local model and global model: $\lambda$.
    
    \State \textbf{Initial:} $w^{Global}\leftarrow{w^{Global}_{0}}$
    \State Dispatch $w^{Global}$ to each client  $ i \in S_k$
    \For{$t = 0, \dots, T-1$}
      \State Generate $\xi = 1$ with a probability $p$
      \If{$\xi$ = 0} 
        \For{each client  $ i \in S_k$, client private data $D^{i} $}
            \State  $w^{i}_{t+1}\leftarrow \Call{LocalUpdate}{w^{i}_{t},D^{i}}$
            \EndFor 
      \Else
        \State ${\bar{w}}_{t} \leftarrow \frac{1}{K} \sum_{S_k}{w^{i}_{t}}$
        \For{each client  $ i \in S_k$}
            \State $w_{t+1}^{i} \leftarrow (1 - \frac{\eta \lambda}{Kp} ) w_{t}^{i} + \frac{\eta \lambda}{Kp} \bar{w}_t$
      
        \EndFor
     \EndIf
     \EndFor
        
  \\\hrulefill
  \Function{LocalUpdate}{$w^{i},D^{i}$}
  \For{epoch $e$ in range($E$)} 
  \State $Loss = L(w^i;D^{i})$
  \State $w^{i}_{t+1} = w^{i}_{t} - \eta \frac{K}{(1-p)} \nabla{Loss}$
  \EndFor
  \State \textbf{Return} $w^{i}_{t+1}$ 
  \EndFunction
  \end{algorithmic}
\end{algorithm}
\section{Experiments}
\subsection{Data Description}
In our work, we use the hourly data from the Pecan Street Inc \cite{pecan-web}. After preprocessing the data and removing incomplete meter readings, we obtained a dataset comprising hourly load data from 44 households over two years. To retrieve the information regarding the weather, we also incorporate weather data, including the temperature and humidity concatenated smart meters' active power as the input data. In order to investigate the impact of weather factors on the performance of disaggregating different appliance loads, we selected a total of 12 appliances, including the low-power appliances. Some of them are intuitively affected by weather (such as air conditioners, pool pumps, and furnaces) and others are likely unrelated to weather (such as light plugs, living room plugs, and bedroom plugs).

To investigate the different performance of three FL frameworks in data heterogeneous and data homogeneous scenarios, We group the data into five clients and generate a heterogeneous dataset and a homogeneous dataset. For the homogeneous dataset, we implement random sampling without putting back strategy. While for the heterogeneous dataset, we manually select data in such a way that the 5th client includes the majority of data when most appliances are turned off, while other clients adopt random sampling methods. Each client only contains one part of the full dataset and they are not allowed to share their private data with other clients and servers during the training. Each client holds 20\% of the full dataset and 70\%, 30\% of clients' data is used for training and testing separately.

\subsection{Model Configurations}
In terms of the sequence-to-sequence model, we choose the bidirectional stacked LSTM both in the encoder and decoder. The layer number is set to five. The hidden units of the encoder and decoder for each layer are set to 128. To do the comparative study, we compare our proposed model with other RNN-based models that are commonly used in the NILM tasks, including the BiLSTM with attention and the Gated Recurrent Unit with attention. The hyperparameters setting for those RNN-based models are the same as the Sequence-to-Sequence model. To inspect the influence of weather factors on the model performance in our load disaggregation task, we do a comparative study that adds weather factors and removes weather factors during the training processing.

For the federated learning, we compare the FedAvg, FedProx and L2GD frameworks on our proposed Seq2Seq model with the addition of the weather factor. According to \cite{li2020federated}, FedProx has the ability to handle the problem caused by data heterogeneity, but more communication roads are required to meet the converge conditions. We compare those three FL frameworks 
in the data heterogeneous and data homogeneous scenarios according to performance and communication efficiency. For the FedProx, we do several experiments to select the hyperparameters $\mu$ and finally choose $\mu = 1e^{-4}$ as the weight of the proximal term in FedProx. For the L2GD, we set hyperparameters $P = 0.33$ and $\lambda = 825$ to perform more local update steps and fewer aggregation steps.
\subsection{Numerical Results and Discussion}\label{sec:discussion}
In this section, We will compare the performance improvement of weather features under center training conditions. Additionally, we will compare the performance and communication efficiency of the L2GD framework under FL training conditions with two other FL frameworks. Furthermore, we will contrast the results of FL and center training to understand the performance trade-offs of using FL training while ensuring data non-sharing.

The comparison configuration are listed below: 
\begin{itemize}

    \item Performance of centralized-trained model (including BiLSTM, GRU, and proposed Seq2Seq model) with and without adding weather features,
    \item Performance and efficiency of Seq2Seq model adding weather features trained in FedAvg, FedProx and L2GD in the homogeneous dataset,
    \item Performance and efficiency of Seq2Seq model adding weather features trained in FedAvg, FedProx and L2GD in the heterogeneous dataset.
    
\end{itemize}

\subsection{Performance Metrics}
To evaluate the performance of load disaggregation results, we introduce four metrics commonly used in the NILM task, including 
\begin{itemize}
    \item Mean-Absolute Error (MAE),
    \item Root-Mean-Square Error (RMSE),
    \item Estimation Accuracy (Eacc) \cite{kolter2011redd},
    \item Normalized Disaggregation Error (NDE) \cite{NDE}.
\end{itemize}

\subsection{Comparison of Federated learning frameworks in homogeneous and heterogeneous datasets}
In this section, we compare the performance and efficiency of the FedAvg, FedProx, and L2GD frameworks in both homogeneous and heterogeneous data scenarios, incorporating weather features and utilizing Seq2Seq models. Table. \ref{tab:heterogeneous dataset result} and Table. \ref{tab:homogeneous result} shows the results of five clients in the heterogeneous and homogeneous datasets respectively. To ensure that the model receives sufficient training without overfitting, we conducted experiments by setting up candidate communication rounds and identified the optimal communication rounds (20 rounds).
As shown in Table.\ref{tab:homogeneous result}, under the conditions of homogeneous data and 20 communication rounds, L2GD, FedProx, and FedAvg exhibit very similar performance across five clients, with Eacc achieving excellent results exceeding 90\%. However, under the conditions of heterogeneous data, as indicated in Table \ref{tab:heterogeneous dataset result}, L2GD outperforms both FedAvg and FedProx on all five clients. This improvement is particularly evident in client 5, where we manually selected low-power consumption data, making the model training more challenging for this client. Significantly, at the same number of communication rounds, L2GD, compared to FedAvg and FedProx, achieves better results on client 5, which has heterogeneous data, without sacrificing the performance of clients with homogeneous data. (Eacc improves by 2.5\% compared to FedAvg and by 2.4\% compared to FedProx. NDE decreases by 21.5\% compared to FedAvg and by 22.1\% compared to FedProx). Since L2GD does not pursue a uniform global model but rather strikes a trade-off between the global model and local models, it allows each client's model to extract features more suited to its data distribution while ensuring a certain level of generalization. This, in turn, leads to better results in heterogeneous scenarios.



\begin{table*}[htbp]
\centering
\caption{Comparison between FL frameworks in the heterogeneous dataset (20 communication rounds)}
\label{tab:heterogeneous dataset result}
\resizebox{\textwidth}{15mm}{
\begin{tabular}{|c|c|c|c|c|c|c|c|c|c|c|c|c|}
\hline
\multirow{2}{*}{\diagbox{\textbf{Clients}}{\textbf{Frameworks}}} & \multicolumn{3}{c|}{\textbf{EACC} }& \multicolumn{3}{c|}{\textbf{RMSE} } & \multicolumn{3}{c|}{\textbf{MAE} } &\multicolumn{3}{c|}{\textbf{NDE}} \\ \cline{2-13} 
 & \textbf{FedAvg} & \textbf{FedProx} &\textbf{L2GD}  & \textbf{FedAvg} & \textbf{FedProx} & \textbf{L2GD} &\textbf{FedAvg}  & \textbf{FedProx} & \textbf{L2GD} & \textbf{FedAvg} & \textbf{FedProx}  &\textbf{L2GD}  \\
\hline
Client 1 & 0.9282& 0.9309& \textbf{0.9298}& 0.0921& 0.0928& \textbf{0.0906}& 0.0224& \textbf{0.0216}& 0.0219& 0.0265& 0.027&\textbf{0.0258}\\ \hline 
Client 2 & 0.8592& 0.8602& \textbf{0.8638}& \textbf{0.1133}& 0.1151& 0.115& 0.0215& 0.0212& \textbf{0.0207}& \textbf{0.0687}& 0.0711&0.0712\\ \hline 
Client 3 & 0.9162& 0.9167& \textbf{0.9229}& 0.0968& 0.0997& \textbf{0.0869}& 0.0211& 0.021& \textbf{0.0194}& 0.0415& 0.0441&\textbf{0.0335}\\ \hline 
Client 4 & 0.8311& 0.8299& \textbf{0.8346}& \textbf{0.099}& 0.1049& 0.1028& 0.0229& 0.0231& \textbf{0.0224}& \textbf{0.0692}& 0.0778&0.0742\\ \hline 
Client 5 & 0.7438& 0.7443& \textbf{0.7626}& 0.0249& 0.0249& \textbf{0.0226}& 0.0066& 0.0065& \textbf{0.0061}& 0.1136& 0.1142&\textbf{0.0935}\\ \hline 

\textbf{Average}& 0.8557& 0.8564& \textbf{0.8627}& 0.0852& 0.0875& \textbf{0.0836}& 0.0189& 0.0187& \textbf{0.0181}& 0.0639& 0.0668&\textbf{0.0596}\\ \hline
\multicolumn{13}{l}{The best results for each clients with different metrics are shown in bold. The client 5 contains more data on appliance in off-state compared to other clients.}

\end{tabular}}
\end{table*}


\begin{table*}[htbp]
\centering
\caption{Comparison between FL frameworks in the homogeneous dataset (20 communication rounds)}
\label{tab:homogeneous result}
\resizebox{\textwidth}{15mm}{
\begin{tabular}{|c|c|c|c|c|c|c|c|c|c|c|c|c|}
\hline
\multirow{2}{*}{\diagbox{\textbf{Clients}}{\textbf{Frameworks}}} & \multicolumn{3}{c|}{\textbf{EACC} }& \multicolumn{3}{c|}{\textbf{RMSE} } & \multicolumn{3}{c|}{\textbf{MAE} } &\multicolumn{3}{c|}{\textbf{NDE}} \\ \cline{2-13} 
 & \textbf{FedAvg} & \textbf{FedProx} &\textbf{L2GD}  & \textbf{FedAvg} & \textbf{FedProx} & \textbf{L2GD} &\textbf{FedAvg}  & \textbf{FedProx} & \textbf{L2GD} & \textbf{FedAvg} & \textbf{FedProx}  &\textbf{L2GD}  \\
\hline
Client 1 & 0.8895& 0.8912& 0.8839& 0.0718& 0.0705& 0.0763& 0.0128& 0.0125& 0.0134& 0.0475& 0.0459&0.0537\\ \hline 
Client 2 & 0.9363& 0.9369& 0.9332& 0.0822& 0.0821& 0.0905& 0.017& 0.0168& 0.0178& 0.0249& 0.025&0.0303\\ \hline 
Client 3 & 0.9029& 0.8992& 0.8949& 0.0655& 0.0717& 0.0716& 0.0116& 0.012& 0.0125& 0.0341& 0.0412&0.0409\\ \hline 
Client 4 & 0.9176& 0.9188& 0.9121& 0.0813& 0.0805& 0.0873& 0.014& 0.0138& 0.0149& 0.038& 0.0373&0.0439\\ \hline 
Client 5 & 0.9202& 0.9218& 0.9188& 0.0772& 0.075& 0.0775& 0.0156& 0.0153& 0.0159& 0.0322& 0.0302&0.0322\\ \hline 

\textbf{Average}& 0.9133& 0.9136& 0.9086& 0.0756& 0.076& 0.0806& 0.0142& 0.0141& 0.0149& 0.0353& 0.0359&0.0402\\ \hline
\end{tabular}}
\end{table*}

To compare the communication efficiency of FedAvg, FedProx, and L2GD, in Table.\ref{tab:homogeneous result 5} and Table.\ref{tab:heterogeneous result 5}, we contrasted the results of the three frameworks with only five communication rounds on both homogeneous and heterogeneous datasets. It is evident that even with only five communication rounds, L2GD outperforms both FedAvg and FedProx, whether on homogeneous or heterogeneous datasets. Most notably, in the case of heterogeneous data, the performance of FedAvg and FedProx on client 5 was significantly deducted, showing underfitted results (Eacc decreases from 0.743 and 0.744 to 0.669 and 0.685, respectively). While L2GD also experiences a decline in performance with reduced communication rounds, the magnitude is smaller. In the case of homogeneous data, when reducing communication rounds, FedAvg and FedProx exhibit a noticeable drop in performance across all clients. However, L2GD, with only five communication rounds, does not show the underfitted issue observed in FedAvg and FedProx. Its performance with five communication rounds is very close to that with twenty communication rounds.

\begin{figure}[ht]
    \centering 
    \includegraphics[width=1.0\linewidth]{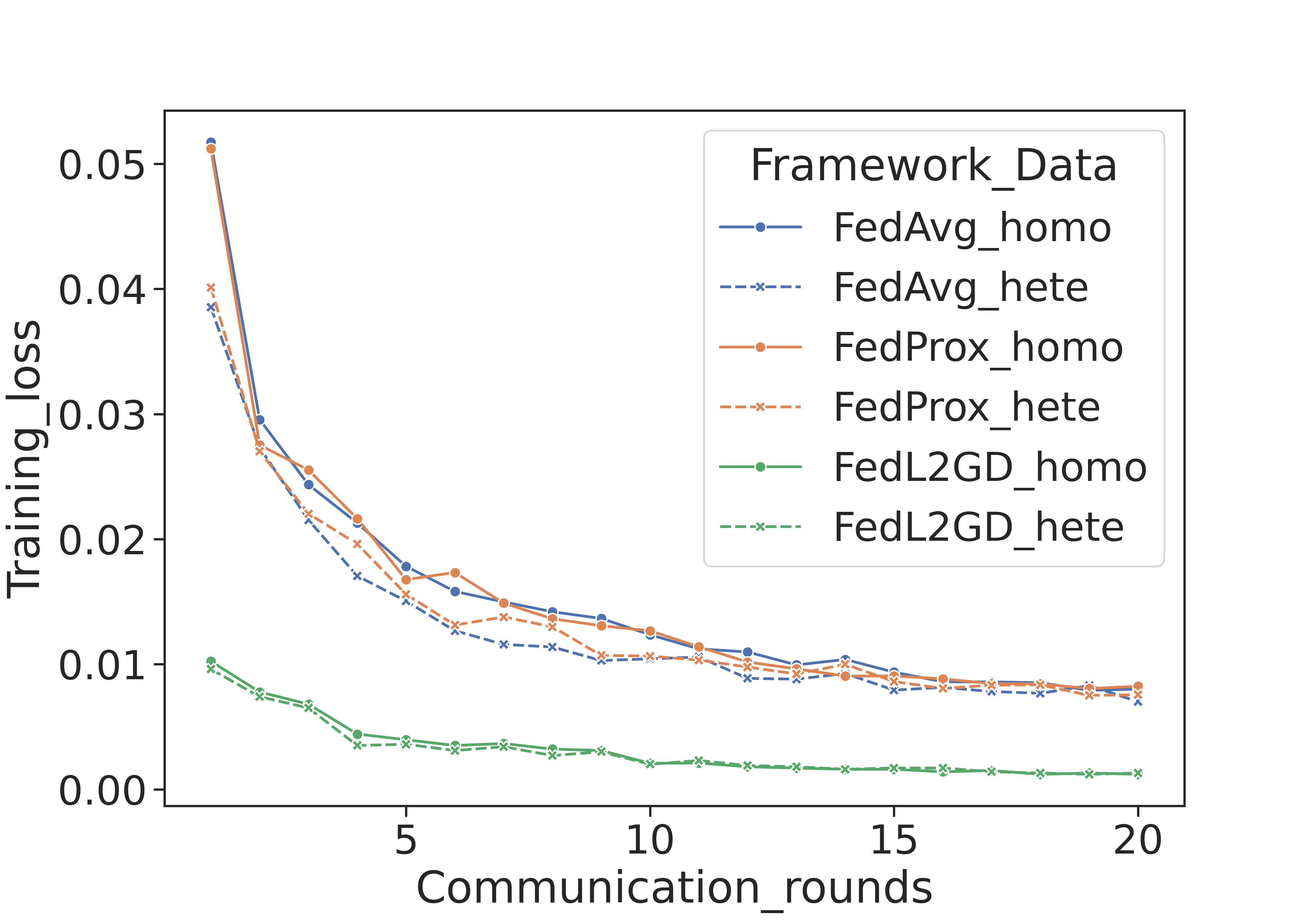} 
    \caption{Average training loss with communication rounds increasing.}
    \label{fig:loss}
\end{figure}

Fig.\ref{fig:loss} shows the average training loss of three FL frameworks as the number of communication rounds increased. The solid line indicates the model trained on the homogeneous dataset while the dash line indicates the model trained on the heterogeneous dataset. When it comes to the FedAvg and FedProx framework, it can be seen that the training loss reduction rates of FedAvg and FedProx are very close. After 5 communication rounds, the MAE of both frameworks decreased from 0.05 to around 0.015, gradually converging to 0.01 thereafter. In the heterogeneous scenario, due to the inclusion of a proximal term as a penalty, FedProx exhibits a slightly higher training loss compared to FedAvg. However, with an increase in communication rounds, the losses of both frameworks gradually converge to the same level. Additionally, in heterogeneous scenarios, where clients overly fit specific data, the training loss reduction is faster for both approaches compared to the homogeneous scenario. For L2GD, we set the aggregation probability P to 0.3. This implies that clients have a higher probability of doing multiple local training iterations before aggregation. As a result, in both heterogeneous and homogeneous scenarios, L2GD requires fewer communication rounds compared to FedAvg and FedProx, enabling faster convergence.

Furthermore, in previous studies, FedProx was considered to mitigate the problems caused by data heterogeneity compared to FedAvg. However, in our task, when the number of communication rounds is 20, there is not a significant difference between FedProx and FedAvg on heterogeneous datasets. However, with a reduced number of communication rounds, FedProx shows a slight improvement over FedAvg on heterogeneous datasets, approaching its performance on homogeneous datasets. We speculate that in heterogeneous scenarios, FedProx can mitigate the impact of heterogeneous data on the convergence speed of the global model, thus achieving better results with limited communication. As for L2GD, since we can adjust the probability of local updates and global aggregation through the hyperparameter P, L2GD can efficiently meet the requirements of both heterogeneous and homogeneous scenarios with limited communication resources.


\begin{table*}[htbp]
\centering
\caption{Comparison between FL frameworks in the heterogeneous dataset (5 communication rounds)}
\label{tab:heterogeneous result 5}
\resizebox{\textwidth}{15mm}{
\begin{tabular}{|c|c|c|c|c|c|c|c|c|c|c|c|c|}
\hline
\multirow{2}{*}{\diagbox{\textbf{Clients}}{\textbf{Frameworks}}} & \multicolumn{3}{c|}{\textbf{EACC} }& \multicolumn{3}{c|}{\textbf{RMSE} } & \multicolumn{3}{c|}{\textbf{MAE} } &\multicolumn{3}{c|}{\textbf{NDE}} \\ \cline{2-13} 
 & \textbf{FedAvg} & \textbf{FedProx} &\textbf{L2GD}  & \textbf{FedAvg} & \textbf{FedProx} & \textbf{L2GD} &\textbf{FedAvg}  & \textbf{FedProx} & \textbf{L2GD} & \textbf{FedAvg} & \textbf{FedProx}  &\textbf{L2GD}  \\
\hline
Client 1 & 0.8949& 0.894& \textbf{0.9096}& 0.1161& 0.1215& \textbf{0.1123}& 0.0328& 0.0331& \textbf{0.0282}& 0.0421& 0.0463&\textbf{0.0393}\\ \hline 
Client 2 & 0.8052& 0.8138& \textbf{0.8335}& 0.1522& 0.1487& \textbf{0.1391}& 0.0296& 0.0283& \textbf{0.0253}& 0.1247& 0.1192&\textbf{0.1033}\\ \hline 
Client 3 & 0.8821& 0.8788& \textbf{0.9014}& 0.1198& 0.1222& \textbf{0.1083}& 0.0297& 0.0305& \textbf{0.0248}& 0.0635& 0.0657&\textbf{0.0519}\\ \hline 
Client 4 & 0.7894& 0.8001& \textbf{0.816}& 0.1208& 0.1093& \textbf{0.106}& 0.0286& 0.0271& \textbf{0.025}& 0.1027& 0.0839&\textbf{0.0789}\\ \hline 
Client 5 & 0.6698& 0.6852& \textbf{0.7201}& 0.0302& 0.0277& \textbf{0.0253}& 0.0084& 0.008& \textbf{0.0072}& 0.1683& 0.1415&\textbf{0.1176}\\ \hline 

\textbf{Average}& 0.8083& 0.8144& \textbf{0.8361}& 0.1078& 0.1059& \textbf{0.0982}& 0.0258& 0.0254& \textbf{0.0221}& 0.1003& 0.0913&\textbf{0.0782}\\ \hline
\multicolumn{13}{l}{The best results for each clients with different metrics are shown in bold. The client 5 contains more data on appliance in off-state compared to other clients.} \\

\end{tabular}}
\end{table*}



\begin{table*}[htbp]
\centering
\caption{Comparison between FL frameworks in homogeneous datase (5 communication rounds)}
\label{tab:homogeneous result 5}
\resizebox{\textwidth}{15mm}{
\begin{tabular}{|c|c|c|c|c|c|c|c|c|c|c|c|c|}
\hline
\multirow{2}{*}{\diagbox{\textbf{Clients}}{\textbf{Frameworks}}} & \multicolumn{3}{c|}{\textbf{EACC} }& \multicolumn{3}{c|}{\textbf{RMSE} } & \multicolumn{3}{c|}{\textbf{MAE} } &\multicolumn{3}{c|}{\textbf{NDE}} \\ \cline{2-13} 
 & \textbf{FedAvg} & \textbf{FedProx} &\textbf{L2GD}  & \textbf{FedAvg} & \textbf{FedProx} & \textbf{L2GD} &\textbf{FedAvg}  & \textbf{FedProx} & \textbf{L2GD} & \textbf{FedAvg} & \textbf{FedProx}  &\textbf{L2GD}  \\
\hline
Client 1 & 0.8399& 0.8449& \textbf{0.8841}& 0.0878& 0.0871& \textbf{0.0732}& 0.0184& 0.0179& \textbf{0.0134}& 0.0715& 0.0701&\textbf{0.0497}\\ \hline 
Client 2 & 0.915& 0.9173& \textbf{0.9335}& 0.0942& 0.096& \textbf{0.0873}& 0.0226& 0.0221& \textbf{0.0177}& 0.033& 0.0337&\textbf{0.0283}\\ \hline 
Client 3 & 0.8562& 0.8583& \textbf{0.8943}& 0.0795& 0.0819& \textbf{0.0704}& 0.0171& 0.0169& \textbf{0.0126}& 0.05& 0.0534&\textbf{0.0394}\\ \hline 
Client 4 & 0.8862& 0.8891& \textbf{0.9099}& 0.0906& 0.092& \textbf{0.0857}& 0.0193& 0.0188& \textbf{0.0153}& 0.0476& 0.0489&\textbf{0.0422}\\ \hline 
Client 5 & 0.8918& 0.8961& \textbf{0.9168}& 0.0865& 0.0857& \textbf{0.0755}& 0.0212& 0.0204& \textbf{0.0163}& 0.0402& 0.0393&\textbf{0.0305}\\ \hline 

\textbf{Average}& 0.8778& 0.8812& \textbf{0.9077}& 0.0877& 0.0885& \textbf{0.0784}& 0.0197& 0.0192& \textbf{0.0151}& 0.0485& 0.0491&\textbf{0.038}\\ \hline
\multicolumn{13}{l}{The best results for each clients with different metrics are shown in bold.}

\end{tabular}}
\end{table*}

\subsection{Effectiveness of Weather Feature to Time-Series Model}

\newcolumntype{a}{>{\columncolor{Gray}}c}

\begin{table*}[htbp]
\caption{Disaggregation results of time-series models that with/without weather features}
    \centering
    \label{tab:weather table}
    \resizebox{\textwidth}{25mm}{
    \begin{tabular}{|c|c|c|a|a|c|c|a|a|c|c|a|a|}
    \hline
    \multirow{2}{*}{\diagbox{\textbf{Appliances}}{\textbf{Models}}} & \multicolumn{4}{c|}{\textbf{BiLSTM with attention \cite{piccialli2021improving}}}& \multicolumn{4}{c|}{\textbf{GRU with attention\cite{jung2021attention}}}& \multicolumn{4}{c|}{\textbf{Seq2Seq (Proposed)}}\\ \cline{2-13}
& \multicolumn{2}{c|}{\textbf{Without weather}}& \multicolumn{2}{a|}{\textbf{With weather}}& \multicolumn{2}{c|}{\textbf{Without weather}}& \multicolumn{2}{a|}{\textbf{With weather}}& \multicolumn{2}{c|}{\textbf{Without weather}}& \multicolumn{2}{a|}{\textbf{With weather}}\\ \cline{2-13}
    
     & \textbf{EACC} & \textbf{NDE} &\textbf{EACC} & \textbf{NDE} & \textbf{EACC} & \textbf{NDE} &\textbf{EACC} & \textbf{NDE} & \textbf{EACC} & \textbf{NDE} &\textbf{EACC} & \textbf{NDE} \\
    \hline
    Ac & 0.9385& 0.0317& 0.9463& 0.0255& 0.9326& 0.0355& 0.9411& 0.0266& 0.9401& 0.0322& 0.9489&0.0246\\ \hline 
    Car & 0.944& 0.027& 0.9486& 0.0237& 0.9384& 0.0297& 0.9429& 0.0258& 0.945& 0.027& 0.9522&0.0216\\ \hline 
    Furnace & 0.8916& 0.1398& 0.9037& 0.1054& 0.8835& 0.149& 0.8986& 0.1133& 0.8945& 0.1414& 0.9106&0.1106\\ \hline 
    Refrigerator & 0.8546& 0.145& 0.8645& 0.1183& 0.8488& 0.1497& 0.8626& 0.1196& 0.8578& 0.1396& 0.8737&0.1099\\ \hline 
    Poolpump & 0.9148& 0.0885& 0.9273& 0.0594& 0.905& 0.0947& 0.918& 0.0755& 0.9098& 0.0982& 0.9309&0.0604\\ \hline 
    Bathroom & 0.8466& 0.1239& 0.8775& 0.0515& 0.8426& 0.1346& 0.8514& 0.1052& 0.8376& 0.1138& 0.8654&0.1059\\ \hline
    Kitchen & 0.7877& 0.1406& 0.8144& 0.0974& 0.7879& 0.1161& 0.7966& 0.1029& 0.7879& 0.1486& 0.8311&0.0978\\  \hline
    Livingroom & 0.7221& 0.4004& 0.7451& 0.328& 0.7048& 0.4241& 0.732& 0.3399& 0.7269& 0.3779& 0.7626&0.3141\\ \hline 
    Bedroom & 0.7249& 0.3706& 0.7494& 0.3504& 0.7091& 0.416& 0.7347& 0.3155& 0.7196& 0.3603& 0.7656&0.335\\ \hline 
    Garage & 0.7196& 0.3239& 0.719& 0.3168& 0.7193& 0.3245& 0.7177& 0.2958& 0.7207& 0.3369& 0.7615&0.2871\\ \hline 
    Office & 0.7747& 0.2773& 0.7955& 0.2294& 0.7568& 0.3242& 0.7719& 0.2609& 0.7709& 0.2956& 0.8011&0.2414\\ \hline 
    Lights\_plugs & 0.7765& 0.2595& 0.7955& 0.2138& 0.766& 0.2695& 0.7873& 0.2285& 0.7868& 0.2376& 0.81&0.2038\\ \hline
    \textbf{Overall}& 0.9085& 2.3282& 0.9178& 1.9197& 0.9017& 2.4676& 0.9116& 2.0095& 0.91& 2.309& 0.9228&1.9122\\ \hline
\multicolumn{13}{l}{ The results with the gray backgrounds
shows the model trained with weather data, while the
white shows the results that without weather features.} 
    
    \end{tabular}}
    \end{table*}

Tabel. \ref{tab:weather table} shows disaggregation results performed by three time-series models. The results with the white backgrounds shows the model trained without weather data, while the gray shows the results that adding weather features. It can be seen that whether the weather feature is included or not, the Seq2Seq model outperforms both the BiLSTM with attention and GRU with attention models. However, without the inclusion of the weather feature, although Seq2Seq is slightly better than the other two models overall, the improvement is not very pronounced for specific appliances. Upon adding the weather feature, a significant enhancement in the performance of all three time-series models can be observed. 
Because the encoder extracts the global information regarding the total load and weather, the Seq2Seq model shows a better performance in low-consumption appliances, the improvement is most notable after incorporating the weather feature.

Additionally, it's interesting to note that in the results of Seq2Seq, appliances more likely to be affected by weather, such as Pool pumps, and Refrigerators, demonstrated improved performance when weather features were incorporated, leading to the reduction of the NDE (reducing 27\% and 62\% respectively). Additionally, even appliances we initially considered less influenced by weather, Living room plugs, Bathroom plugs, and Kitchen plugs, showed significant performance enhancements with the inclusion of weather features (the Eacc improved 4.9\%, 3.3\%, 5.5\% respectively). The performance of Bathroom plugs and Kitchen plugs improved significantly, possibly due to residents' electricity usage habits being more sensitive to weather conditions (e.g., increased usage of electric hairdryers during winter).

\section{Conclusion and Future Work}
In this paper, we introduce a federated sequence-to-sequence load disaggregation method by using the hourly sampled active power data and easily accessible weather data to perform load disaggregation tasks both for low-consumption and high-consumption appliances. Additionally, We have, for the first time in the NILM domain, deployed the L2GD framework in conjunction with our proposed Seq2Seq model. This implementation achieves efficient communication for load disaggregation while preserving data privacy. The experimental results indicate that L2GD not only enables efficient communication but also outperforms FedAvg and FedProx, particularly in NILM tasks‘ of heterogeneous scenarios.

In future work, we will focus on identifying other easily accessible features, similar to weather data, to facilitate load disaggregation tasks in scenarios with very low sampling rates.



\bibliographystyle{IEEEtran}
\bibliography{bibfile}

\begin{thebibliography}{10}
\providecommand{\url}[1]{#1}
\csname url@samestyle\endcsname
\providecommand{\newblock}{\relax}
\providecommand{\bibinfo}[2]{#2}
\providecommand{\BIBentrySTDinterwordspacing}{\spaceskip=0pt\relax}
\providecommand{\BIBentryALTinterwordstretchfactor}{4}
\providecommand{\BIBentryALTinterwordspacing}{\spaceskip=\fontdimen2\font plus
\BIBentryALTinterwordstretchfactor\fontdimen3\font minus \fontdimen4\font\relax}
\providecommand{\BIBforeignlanguage}[2]{{%
\expandafter\ifx\csname l@#1\endcsname\relax
\typeout{** WARNING: IEEEtran.bst: No hyphenation pattern has been}%
\typeout{** loaded for the language `#1'. Using the pattern for}%
\typeout{** the default language instead.}%
\else
\language=\csname l@#1\endcsname
\fi
#2}}
\providecommand{\BIBdecl}{\relax}
\BIBdecl

\bibitem{benzi2011electricity}
F.~Benzi, N.~Anglani, E.~Bassi, and L.~Frosini, ``Electricity smart meters interfacing the households,'' \emph{IEEE Transactions on Industrial Electronics}, vol.~58, no.~10, pp. 4487--4494, 2011.

\bibitem{hart1992nonintrusive}
G.~W. Hart, ``Nonintrusive appliance load monitoring,'' \emph{Proceedings of the IEEE}, vol.~80, no.~12, pp. 1870--1891, 1992.

\bibitem{eb2018}
A.~U. Rehman, T.~T. Lie, B.~Valles, and S.~R. Tito, ``Low complexity event detection algorithm for non- intrusive load monitoring systems,'' in \emph{2018 IEEE Innovative Smart Grid Technologies - Asia (ISGT Asia)}, 2018, pp. 746--751.

\bibitem{eb2020}
A.~U. Rehman, T.~T. Lie, B.~Vallès, and S.~R. Tito, ``Event-detection algorithms for low sampling nonintrusive load monitoring systems based on low complexity statistical features,'' \emph{IEEE Transactions on Instrumentation and Measurement}, vol.~69, no.~3, pp. 751--759, 2020.

\bibitem{7007702}
J.~Liao, G.~Elafoudi, L.~Stankovic, and V.~Stankovic, ``Non-intrusive appliance load monitoring using low-resolution smart meter data,'' in \emph{2014 IEEE International Conference on Smart Grid Communications (SmartGridComm)}, 2014, pp. 535--540.

\bibitem{stankovic2016measuring}
L.~Stankovic, V.~Stankovic, J.~Liao, and C.~Wilson, ``Measuring the energy intensity of domestic activities from smart meter data,'' \emph{Applied Energy}, vol. 183, pp. 1565--1580, 2016.

\bibitem{CNNmulti}
\BIBentryALTinterwordspacing
A.~Faustine, L.~Pereira, H.~Bousbiat, and S.~Kulkarni, ``Unet-nilm: A deep neural network for multi-tasks appliances state detection and power estimation in nilm,'' in \emph{Proceedings of the 5th International Workshop on Non-Intrusive Load Monitoring}, ser. NILM'20.\hskip 1em plus 0.5em minus 0.4em\relax New York, NY, USA: Association for Computing Machinery, 2020, p. 84–88. [Online]. Available: \url{https://doi.org/10.1145/3427771.3427859}
\BIBentrySTDinterwordspacing

\bibitem{zhou2020sequence}
G.~Zhou, Z.~Li, M.~Fu, Y.~Feng, X.~Wang, and C.~Huang, ``Sequence-to-sequence load disaggregation using multiscale residual neural network,'' \emph{IEEE Transactions on Instrumentation and Measurement}, vol.~70, pp. 1--10, 2020.

\bibitem{CNNBilstm}
\BIBentryALTinterwordspacing
M.~Kaselimi and N.~Doulamis, ``Long-term recurrent convolutional networks for non-intrusive load monitoring,'' in \emph{Proceedings of the 13th ACM International Conference on PErvasive Technologies Related to Assistive Environments}, ser. PETRA '20.\hskip 1em plus 0.5em minus 0.4em\relax New York, NY, USA: Association for Computing Machinery, 2020. [Online]. Available: \url{https://doi.org/10.1145/3389189.3397995}
\BIBentrySTDinterwordspacing

\bibitem{kelly2015neural}
\BIBentryALTinterwordspacing
J.~Kelly and W.~Knottenbelt, ``Neural nilm: Deep neural networks applied to energy disaggregation,'' in \emph{Proceedings of the 2nd ACM International Conference on Embedded Systems for Energy-Efficient Built Environments}, ser. BuildSys '15.\hskip 1em plus 0.5em minus 0.4em\relax New York, NY, USA: Association for Computing Machinery, 2015, p. 55–64. [Online]. Available: \url{https://doi.org/10.1145/2821650.2821672}
\BIBentrySTDinterwordspacing

\bibitem{wang2019new}
T.~Wang, T.~Ji, and M.~Li, ``A new approach for supervised power disaggregation by using a denoising autoencoder and recurrent lstm network,'' in \emph{2019 IEEE 12th International symposium on diagnostics for electrical machines, power electronics and drives (SDEMPED)}.\hskip 1em plus 0.5em minus 0.4em\relax Toulouse, France: IEEE, 2019, pp. 507--512.

\bibitem{basu2015time}
K.~Basu, V.~Debusschere, A.~Douzal-Chouakria, and S.~Bacha, ``Time series distance-based methods for non-intrusive load monitoring in residential buildings,'' \emph{Energy and Buildings}, vol.~96, pp. 109--117, 2015.

\bibitem{7793294}
K.~Basu, A.~Hably, V.~Debusschere, S.~Bacha, G.~J. Driven, and A.~Ovalle, ``A comparative study of low sampling non intrusive load dis-aggregation,'' in \emph{IECON 2016 - 42nd Annual Conference of the IEEE Industrial Electronics Society}, 2016, pp. 5137--5142.

\bibitem{6915905}
K.~Basu, V.~Debusschere, S.~Bacha, U.~Maulik, and S.~Bondyopadhyay, ``Nonintrusive load monitoring: A temporal multilabel classification approach,'' \emph{IEEE Transactions on Industrial Informatics}, vol.~11, no.~1, pp. 262--270, 2015.

\bibitem{quy2021data}
T.~L. Quy, S.~Zerr, E.~Ntoutsi, and W.~Nejdl, ``Data augmentation for dealing with low sampling rates in nilm,'' \emph{arXiv preprint arXiv:2104.02055}, 2021.

\bibitem{fagiani2019non}
M.~Fagiani, R.~Bonfigli, E.~Principi, S.~Squartini, and L.~Mandolini, ``A non-intrusive load monitoring algorithm based on non-uniform sampling of power data and deep neural networks,'' \emph{Energies}, vol.~12, no.~7, p. 1371, 2019.

\bibitem{kipping2016modeling}
A.~Kipping and E.~Tr{\o}mborg, ``Modeling and disaggregating hourly electricity consumption in norwegian dwellings based on smart meter data,'' \emph{Energy and Buildings}, vol. 118, pp. 350--369, 2016.

\bibitem{birt2012disaggregating}
B.~J. Birt, G.~R. Newsham, I.~Beausoleil-Morrison, M.~M. Armstrong, N.~Saldanha, and I.~H. Rowlands, ``Disaggregating categories of electrical energy end-use from whole-house hourly data,'' \emph{Energy and buildings}, vol.~50, pp. 93--102, 2012.

\bibitem{zhang2015residential}
G.~Zhang, G.~Wang, H.~Farhangi, and A.~Palizban, ``Residential electric load disaggregation for low-frequency utility applications,'' in \emph{2015 IEEE Power \& Energy Society General Meeting}.\hskip 1em plus 0.5em minus 0.4em\relax Denver, CO, USA: IEEE, 2015, pp. 1--5.

\bibitem{giuseppi2022decentralized}
A.~Giuseppi, S.~Manfredi, D.~Menegatti, A.~Pietrabissa, and C.~Poli, ``Decentralized federated learning for nonintrusive load monitoring in smart energy communities,'' in \emph{2022 30th Mediterranean Conference on Control and Automation (MED)}.\hskip 1em plus 0.5em minus 0.4em\relax Vouliagmeni, Greece: IEEE, 2022, pp. 312--317.

\bibitem{kaspour2022federated}
S.~Kaspour and A.~Yassine, ``A federated learning model with short sequence to point mechanism for smart home energy disaggregation,'' in \emph{2022 IEEE Symposium on Computers and Communications (ISCC)}.\hskip 1em plus 0.5em minus 0.4em\relax Rhodes, Greece: IEEE, 2022, pp. 1--6.

\bibitem{li2020federated}
T.~Li, A.~K. Sahu, M.~Zaheer, M.~Sanjabi, A.~Talwalkar, and V.~Smith, ``Federated optimization in heterogeneous networks,'' \emph{Proceedings of Machine learning and systems}, vol.~2, pp. 429--450, 2020.

\bibitem{de2021recurrent}
L.~de~Diego-Ot{\'o}n, D.~Fuentes-Jimenez, {\'A}.~Hern{\'a}ndez, and R.~Nieto, ``Recurrent lstm architecture for appliance identification in non-intrusive load monitoring,'' in \emph{2021 IEEE International Instrumentation and Measurement Technology Conference (I2MTC)}.\hskip 1em plus 0.5em minus 0.4em\relax IEEE, 2021, pp. 1--6.

\bibitem{mauch2015new}
L.~Mauch and B.~Yang, ``A new approach for supervised power disaggregation by using a deep recurrent lstm network,'' in \emph{2015 IEEE Global Conference on Signal and Information Processing (GlobalSIP)}.\hskip 1em plus 0.5em minus 0.4em\relax IEEE, 2015, pp. 63--67.

\bibitem{kaselimi2019bayesian}
M.~Kaselimi, N.~Doulamis, A.~Doulamis, A.~Voulodimos, and E.~Protopapadakis, ``Bayesian-optimized bidirectional lstm regression model for non-intrusive load monitoring,'' in \emph{ICASSP 2019-2019 IEEE International Conference on Acoustics, Speech and Signal Processing (ICASSP)}.\hskip 1em plus 0.5em minus 0.4em\relax IEEE, 2019, pp. 2747--2751.

\bibitem{zhou2022non}
Y.~Zhou, C.~Ji, Z.~Dong, L.~Yang, and L.~Zhou, ``Non-intrusive load disaggregation based on multiple optimization of appliance features and cnn-nlstm model,'' in \emph{2022 IEEE/IAS Industrial and Commercial Power System Asia (I\&CPS Asia)}.\hskip 1em plus 0.5em minus 0.4em\relax IEEE, 2022, pp. 871--876.

\bibitem{liu2022nonintrusive}
H.~Liu, L.~Li, G.~Ding, and Q.~Zhang, ``Nonintrusive load disaggregation combining with external attention mechanism and seq2piont,'' in \emph{2022 7th International Conference on Power and Renewable Energy (ICPRE)}.\hskip 1em plus 0.5em minus 0.4em\relax IEEE, 2022, pp. 620--624.

\bibitem{fan2022bidirectional}
Y.~Fan, C.~Liu, T.~Guo, and D.~Jiang, ``Bidirectional attention lstm networks for non-instructive load monitoring,'' in \emph{2022 Prognostics and Health Management Conference (PHM-2022 London)}.\hskip 1em plus 0.5em minus 0.4em\relax IEEE, 2022, pp. 399--404.

\bibitem{zhang2021improving}
J.~Zhang, J.~Sun, J.~Gan, Q.~Liu, and X.~Liu, ``Improving domestic nilm using an attention-enabled seq2point learning approach,'' in \emph{2021 IEEE Intl Conf on Dependable, Autonomic and Secure Computing, Intl Conf on Pervasive Intelligence and Computing, Intl Conf on Cloud and Big Data Computing, Intl Conf on Cyber Science and Technology Congress (DASC/PiCom/CBDCom/CyberSciTech)}.\hskip 1em plus 0.5em minus 0.4em\relax IEEE, 2021, pp. 434--439.

\bibitem{hudson2021framework}
N.~Hudson, M.~J. Hossain, M.~Hosseinzadeh, H.~Khamfroush, M.~Rahnamay-Naeini, and N.~Ghani, ``A framework for edge intelligent smart distribution grids via federated learning,'' in \emph{2021 International Conference on Computer Communications and Networks (ICCCN)}.\hskip 1em plus 0.5em minus 0.4em\relax IEEE, 2021, pp. 1--9.

\bibitem{Seq2point_FL}
Y.~Zhang, G.~Tang, Q.~Huang, Y.~Wang, K.~Wu, K.~Yu, and X.~Shao, ``Fednilm: Applying federated learning to nilm applications at the edge,'' \emph{IEEE Transactions on Green Communications and Networking}, pp. 1--1, 2022.

\bibitem{li2021energy}
Q.~Li, J.~Ye, W.~Song, and Z.~Tse, ``Energy disaggregation with federated and transfer learning,'' in \emph{2021 IEEE 7th World Forum on Internet of Things (WF-IoT)}.\hskip 1em plus 0.5em minus 0.4em\relax IEEE, 2021, pp. 698--703.

\bibitem{liu2022learning}
R.~Liu and Y.~Chen, ``Learning task-aware energy disaggregation: a federated approach,'' in \emph{2022 IEEE 61st Conference on Decision and Control (CDC)}.\hskip 1em plus 0.5em minus 0.4em\relax IEEE, 2022, pp. 4412--4418.

\bibitem{hanzely2020federated}
F.~Hanzely and P.~Richt{\'a}rik, ``Federated learning of a mixture of global and local models,'' \emph{arXiv preprint arXiv:2002.05516}, 2020.

\bibitem{luong-etal-2015-effective}
\BIBentryALTinterwordspacing
T.~Luong, H.~Pham, and C.~D. Manning, ``Effective approaches to attention-based neural machine translation,'' in \emph{Proceedings of the 2015 Conference on Empirical Methods in Natural Language Processing}.\hskip 1em plus 0.5em minus 0.4em\relax Lisbon, Portugal: Association for Computational Linguistics, Sep. 2015, pp. 1412--1421. [Online]. Available: \url{https://aclanthology.org/D15-1166}
\BIBentrySTDinterwordspacing

\bibitem{Sparsemax}
A.~F.~T. Martins and R.~F. Astudillo, ``From softmax to sparsemax: A sparse model of attention and multi-label classification,'' in \emph{Proceedings of the 33rd International Conference on International Conference on Machine Learning - Volume 48}, ser. ICML'16.\hskip 1em plus 0.5em minus 0.4em\relax New York, NY, USA: JMLR.org, 2016, p. 1614–1623.

\bibitem{pecan-web}
\BIBentryALTinterwordspacing
{Street Pecan}, ``Pecan street online database,'' [Accessed July 2019]. [Online]. Available: \url{https://www.pecanstreet.org/work/energy/}
\BIBentrySTDinterwordspacing

\bibitem{kolter2011redd}
J.~Z. Kolter and M.~J. Johnson, ``Redd: A public data set for energy disaggregation research,'' in \emph{Workshop on data mining applications in sustainability (SIGKDD), San Diego, CA}, vol.~25.\hskip 1em plus 0.5em minus 0.4em\relax San Diego, CA: Citeseer, 2011, pp. 59--62.

\bibitem{NDE}
P.~B.~M. Martins, J.~G. R.~C. Gomes, V.~B. Nascimento, and A.~R. de~Freitas, ``Application of a deep learning generative model to load disaggregation for industrial machinery power consumption monitoring,'' in \emph{2018 IEEE International Conference on Communications, Control, and Computing Technologies for Smart Grids (SmartGridComm)}, 2018, pp. 1--6.

\bibitem{piccialli2021improving}
V.~Piccialli and A.~M. Sudoso, ``Improving non-intrusive load disaggregation through an attention-based deep neural network,'' \emph{Energies}, vol.~14, no.~4, p. 847, 2021.

\bibitem{jung2021attention}
S.~Jung, J.~Moon, S.~Park, and E.~Hwang, ``An attention-based multilayer gru model for multistep-ahead short-term load forecasting,'' \emph{Sensors}, vol.~21, no.~5, p. 1639, 2021.

\end{thebibliography}

\end{document}